\documentclass[twocolumn,showpacs,preprintnumbers,amsmath,amssymb]{revtex4}
\usepackage[dvips]{graphicx}

\begin{document}
\title{
The effect of random positions for dipole hopping through a Rydberg gas}
\author{F. Robicheaux$^1$}
\email{robichf@purdue.edu}
\author{N. M. Gill$^2$}
\affiliation{$^1$Department of Physics, Purdue University, West Lafayette,
Indiana 47907, USA}
\affiliation{$^2$Department of Physics, Auburn University, AL
36849, USA}
\date{\today}

\begin{abstract}
We calculate the effect of two kinds of randomness on the hopping of
an excitation through a nearly regular Rydberg gas. We
present calculations for how fast the excitation can hop
away from its starting position for different dimensional lattices
and for different levels of randomness. We also examine the asymptotic
in time final position of the excitation to determine whether
or not the excitation can be localized. The one dimensional
system is an example of Anderson localization where the randomness
is in the off-diagonal elements although the long-range nature of the
interaction leads to non-exponential decay with distance. The
two dimensional square lattice shows a mixture of extended and
localized states for large randomness while there is no visible sign
of localized states for weak randomness. The three dimensional cubic
lattice has few localized states even for strong randomness.
\end{abstract}


\maketitle

\section{Introduction}

The interaction of many atoms and/or molecules can lead to a rich
variety of processes. There has been recent interest in the physics
of many atoms interacting with each other through the dipole-dipole
potential. This interest is spurred by the developments in experiments
and calculations of cold gases. There have been studies of amorphous
systems where the atoms/molecules have a random placement as well as studies
of atoms/molecules placed on perfect lattices. The purpose of this
paper is to present results of a system with the atoms placed in
a lattice but with some randomness in the placement.

The system discussed in this paper is a lattice of Rydberg atoms with
dipole-dipole interactions. This system is a more regular arrangement
of atoms but is otherwise similar in spirit to the original experiments
on Rydberg gases.\cite{AVG,MCT} In these experiments, a dense Rydberg
gas was achieved by exciting many atoms to a Rydberg state and
the subsequent dipole-dipole interactions between atoms
caused the state to change; the new states could then hop through
the sea of unchanged states.
We will treat an idealized case of a lattice of Rydberg
atoms where every atom except one is in a highly excited $s$-state
and the exception is a $p$-state. Because of the dipole-dipole
interaction, the $p$-state can hop from atom-to-atom.

Rydberg gases are systems that can display a wide variety of
physical effects. Reference~\cite{JCZ} described calculations that
showed the strongly interacting Rydberg atoms could shift the
energy of the pair out of resonance which provides a blockade
to further excitation. Reference~\cite{TFS} provided experimental
evidence for this effect by showing the number of Rydberg atoms
excited in a dense gas did not scale linearly with the laser power.
Extreme examples of this effect were described in Ref.~\cite{HRB}
where more than 1000 atoms were blockaded and in Refs.~\cite{UJH,GMW}
where the blockade effect was demonstrated between two individual atoms.
Reference~\cite{ABV} provided spectroscopic evidence for the
dipole-dipole interaction between cold Rydberg atoms.
Reference~\cite{PDL} gave results of calculations that showed
an optimal choice in the laser parameters could lead to the Rydberg
atoms being in a regular spatial array even though the ground state
atoms are randomly distributed in a gas. Again, by detuning the
laser excitation of the Rydberg atoms, Ref.~\cite{AGH} gave
experimental evidence for an antiblockade. As a final example of
basic phenomena, Ref.~\cite{DKH} gave experimental and theoretical
evidence for spatially resolved observation of the effect of
dipole-dipole interaction between Rydberg states.

Besides the basic phenomena of interacting Rydberg atoms, there
have been studies of collective effects in these strongly
interacting gases. For example, Ref.~\cite{WLP} gave the results
of calculations of quantum critical behavior and the appearance
of correlated many-body phenomena. Reference~\cite{JAL} computationally
studied a two-dimensional Rydberg gas and its relationship with
the quantum hard-squares model. Calculations of the coupling of
weak light to Rydberg states of atoms suggested the possibility
for Wigner crystals made up of single photons.\cite{OMM}
Experimental results from exciting a two-dimensional Mott insulator
to Rydberg states\cite{SCE} found the emergence of spatially ordered
excitation patterns; semiclassical calculations of this system\cite{DP}
found highly sub-Poissonian distribution of the number of excitations.
As a final example, Ref.~\cite{ZMB} calculated the static properties
of laser excited Rydberg atoms in one- and two-dimensional lattices.

This paper explores the role of randomness in the hopping of a
Rydberg excitation of one type through a sea of Rydberg atoms.
We investigate the case of having the atoms placed on a lattice
of sites.
Reference~\cite{AYR} has successfully trapped Rydberg atoms in an
optical lattice, but this situation could also be created by taking
ground state atoms trapped in an optical lattice and exciting them
to a Rydberg state. Although there have been studies of a lattice
of Rydberg atoms,\cite{RHT} we do not know of any where the role of disorder
is investigated. We have calculated the distribution of hopping
distances as a function of time for one-, two-, and three-dimensional
lattices with varying amounts of disorder. We also investigate the
unphysical $t\to\infty$ limit of the distribution of hopping distances
as a way to determine whether the hopping was slowed by the randomness
or whether the hopping was localized.

The question of randomness in this system allows us to connect to
Anderson localization\cite{PAW} which has been observed in many
systems including light traveling in a dielectric.\cite{WBL}
For short range Hamiltonians, one-dimensional lattices have all
eigenstates localized even for small amounts of disorder. We find that
the hopping due to the dipole-dipole interaction also has every
state localized. However, the distribution of hopping distances
has a stretched exponential decay for small to intermediate hopping
distances and a power law decay for larger distances. For two-
and three-dimensions, we find that almost no states are localized
for weak randomness. Even for strong randomness, only a small
fraction of states are localized. Finally, we note that the dipole-dipole
interaction has the same form for magnetic and electric dipoles.
Thus, although the details may differ, our results are also applicable
to atoms with magnetic dipoles trapped in an optical lattice.

Atomic units are used unless explicit SI units are given.

\section{Computational method}

To obtain specific results, we solved for a particular case of
dipole hopping through a Rydberg gas. We treat the case where
one atom is a $p$-state and all of the other atoms are $s$-states.
The two states should have similar principle quantum number so
that the dipole coupling between states is as large as possible.
For the cases treated in this paper, we chose the $30s$ and $30p$
states of Rb.

\subsection{Hamiltonian}

This special case ($p$-state hopping through a sea of $s$-states)
is treated as Eq.~(6) in Ref.~\cite{RHT}.
The basis states can be labeled as $|i,m\rangle$ meaning the
$p$-state is at site $i$ with angular momentum projection $m$.
In this special case, the non-zero matrix elements reduce to
\begin{eqnarray}
&\null &V_{im,i'm'}=-\sqrt{\frac{8\pi}{3}}\frac{(d_{n_a1,n_b0})^2}{R^3}\cr
&\null&\times(-1)^{m'}
\begin{pmatrix}1 & 1 & 2\cr m & -m'&m'-m\end{pmatrix}
Y_{2,m'-m}(\hat{R})\label{eqpot}
\end{eqnarray}
where the $d_{n_a1,n_b0}$ is the reduced matrix element between
the $p$-state with principle quantum number $n_a$ and the $s$-state
with principle quantum number $n_b$,
$(...)$ is the usual 3-$j$ coefficient and $\vec{R}=\vec{r}_i
-\vec{r}_{i'}$ is the displacement vector between sites $i$ and $i'$.

For the general case, the non-zero matrix elements are complex. In order
to treat the largest number of atoms, we further restricted the $p$-state
to have $m=0$. This can be accomplished experimentally by having an
external field so that the $m=0,1,-1$ states are sufficiently
separated in energy so that the hopping does not mix $m$. Now the
basis state can be designated solely by the site $i$ and the
non-zero matrix elements reduce to
\begin{equation}\label{eqPot}
H_{ii'}=V_{ii'}=-\frac{2}{3}P_2(\cos\theta_{ii'})\frac{(d_{n_a1,n_b0})^2}{R^3}
\end{equation}
where $P_2(x)=(3x^2-1)/2$ is a Legendre polynomial and $\cos\theta_{ii'}
=(z_i-z_{i'})/R$. This expression is only for $i\neq i'$; when
$i=i'$, the matrix element is 0:
$H_{ii}=0$. By choosing $m=0$ for the $p$-state, the Hamiltonian
will be real, symmetric which means the eigenvectors and eigenvalues
will be real; this will reduce the amount of computer memory and time
needed for the calculations.
In all of the calculations, we use wrap boundary conditions in order to
get better estimates of infinite size systems.

\subsection{Randomness}

We performed calculations for two kinds of randomness in the system.

Type (1) randomness has an atom at every site but there is a random
shift of each atom. The $x$-position of the $i$-th atom is shifted from
the perfect placement by an amount $(Ran - 0.5)*\eta*\delta x$ where $Ran$
is a random number with a flat distribution between 0 and 1 and
$\delta x$ is the spacing of atoms in the $x$-direction. There is a
similar randomness introduced into the $y$-position for the 2-
and 3-dimensional calculations. Finally, there is a similar randomness
introduced in the $z$ position for the 3-dimensional calculation.
Thus, the randomness is only in the directions of the lattice position
of the 
atom (in 1-dimension, the randomness is only in the $x$-placement,
etc.). The parameter $\eta$ characterizes the amount of randomness.
When $\eta =0$, the system is a perfect lattice. We performed
calculations for $\eta = 0$, 0.1, 0.2, 0.3, 0.4, and 0.5.

Type (2) randomness has the atoms placed perfectly on a lattice
but each site may or may not be occupied. We did this calculation
by generating a random number for each site. If the random number
was greater than a parameter $\zeta$, then the site would be occupied.
On average, the number of occupied sites is $1-\zeta$ times the
number of sites in the lattice, and, thus, $\zeta$ is the average
fraction of missing sites. To compare with calculations
of Type (1) randomness, we compare cases where the lattice sizes
are the same. We performed calculations for $\zeta = 0$, 0.1, 0.2,
0.3, 0.4, and 0.5.

\subsection{Distribution of hopping distances}

We were interested both in the time dependence of how an excitation
hops through an imperfect lattice and in the asymptotic, $t\to\infty$,
distribution of sites the excitation can reach. If we were only
interested in the time dependence of the hopping, we could compute
the distribution using several different methods for solving the
time dependent Schrodinger equation. However, the
asymptotic distribution can be found simply from the eigenvalues
and eigenvectors of the Hamiltonian; since we needed the eigenvalues
and eigenvectors for the asymptotic calculation, we also used them
for the time dependent calculations.

The amplitude that an excitation starts at site $i$ at $t=0$ and
hops to site $j$ at time $t$ can be found from the eigenvalues
and eigenvectors as
\begin{equation}
A_{j\leftarrow i}(t)=\sum_\alpha U_{j\alpha}e^{-iE_\alpha t}
U^\dagger_{\alpha i}
\end{equation}
where the $U$ and $E_{\alpha}$ are the eigenvectors and eigenvalues
of the Hamiltonian described in the previous section
\begin{equation}
\sum_i H_{ji}U_{i\alpha} = U_{j\alpha} E_\alpha .
\end{equation}
The probability for the excitation to start at site $i$ and hop to
site $j$ at time $t$ is simply
\begin{equation}
P_{j\leftarrow i}(t)=|A_{j\leftarrow i}(t)|^2
\end{equation}
which is the standard definition for probability.

The asymptotic probability to start at site $i$ and be at site $j$
can be defined as
\begin{equation}
P_{j\leftarrow i}(\infty )\equiv \lim_{t\to\infty} \frac{1}{T}\int_t^{t+T}
P_{j\leftarrow i}(t')dt'
\end{equation}
where $T\gg \hbar /\Delta E$ with $\Delta E$ the smallest energy
difference in the system. One can show that this is equivalent to
\begin{equation}\label{eqAsym}
P_{j\leftarrow i}(\infty )=\sum_\alpha |U_{i\alpha}|^2|U_{j\alpha}|^2
\end{equation}

Once we have the probability for an excitation starting at a site $i$ to
be at a site $j$, the calculation of the distribution of hopping
distance can be obtained by binning. The probability to have hopped
to a site a distance between $r$ and $r+\delta r$ is defined
as $D(r)\delta r$. With this definition, the distribution of hopping
distances is given as
\begin{equation}\label{eqDhop}
D(k\delta r,t) =\frac{1}{N\delta r}\sum_{i,j}P_{j\leftarrow i}(t)
\Xi (r_{ij}-k\delta r)
\end{equation}
where $k$ is a non-negative integer,
$r_{ij}=|\vec{r_i}-\vec{r_j}|$ is the distance between sites
$i$ and $j$, and $\Xi (x)$ equals 1 for $0<x<\delta r$ and
is 0 otherwise. Of course, the algorithm we
use to implement this definition involves taking the integer
part of $r_{ij}/\delta r$ to find $k$. In our calculations,
we average the distribution of hopping distances over many
different random configurations to obtain our final results.

For one dimension, the $D(r)$ will be a decreasing function of
$r$ because the excitation will tend to be near its original
position. For higher dimension, the $D(r)$ can exhibit a more
complicated dependence with $r$ because there are more sites
between $r$ and $r+\delta r$ as $r$ increases. In two dimension,
the number of sites between $r$ and $r+\delta r$ increases
linearly with $r$ while in three dimensions it increases
quadratically.

\section{Results}

In all of our calculations, we use the $30s$ and $30p$ states
of Rb as our `sea' and `hopper' states respectively. The size of
these states is less than 0.1~$\mu$m. The standard step distance
between atoms will be 10~$\mu$m. Thus, the interactions higher
order than dipole-dipole should be negligible. These states have
dipole matrix element $d_{30s,30p}=846$~a.u.
For the one- and two-dimensional
calculations, the atoms will be confined in the $xy$-plane which
means there is no angular dependence to the hopping matrix elements.
For the three-dimensional calculation, there is an angular
dependence due to the $\cos^2(\theta_{ii'})$ term in the matrix
element.

A useful quantity is the energy scale of the matrix element
between nearest neighbors:
$E_{sc}\sim d^2/R^3=1.06\times 10^{-10}$~a.u. We can convert
this to a time scale by $t_{sc}=2\pi /E_{sc}=5.93\times 10^{10}$~a.u.
which is 1.43~$\mu$s. This gives a sense of the time scale needed
for the $p$-state character to hop from site to site.

\subsection{One dimension}

In this section, we present the results of our calculations for
a one dimensional lattice. For the Type (1) randomness, the atoms
are only shifted in the $x$-direction.

\subsubsection{Type (1) Randomness}

\begin{figure}
\resizebox{120mm}{!}{\includegraphics{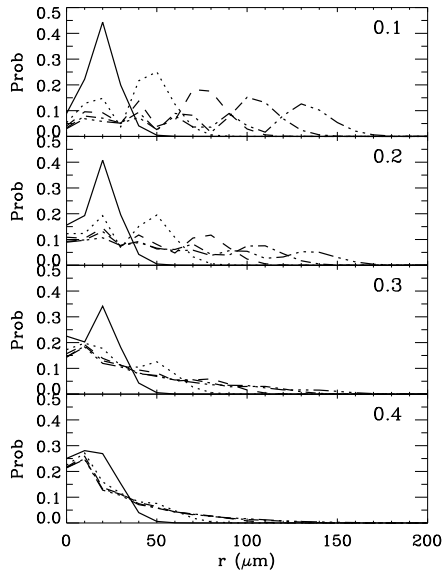}}
\caption{The probability for the $p$-state to be a distance
$r$ from its starting point averaged over a radial region
equal to a lattice spacing. All of these calculations are for
one dimension and have
Type (1) randomness with the top graph having $\eta = 0.1$
and the bottom graph having $\eta = 0.4$. The solid line
is at a time of 1~$\mu$s, the dotted line is at
2~$\mu$s, the dashed line is at 3~$\mu$s, the dot-dash line
is at 4~$\mu$s, and the dash-dot-dot-dot line is at
5~$\mu$s. Since the perfect lattice spacing is 10~$\mu$m,
the graphs show the region within 20 lattice spacings.
}
\end{figure}

Figure~1 gives an indication of how the $p$-state moves away
from the atom it starts on. This figure shows the distribution
of hopping distances, Eq.~(\ref{eqDhop}), times the lattice
spacing, 10~$\mu$m. Thus, the $y$-axis is the probability to
find the $p$-state on a lattice site a distance $r$ from
its initial position. The distribution for times of 1, 2, 3,
4, and 5 $\mu$s are shown for different levels of Type (1)
randomness. We do not show the calculations for $\eta = 0$
(a perfect lattice) because it hardly differs from the
$\eta = 0.1$ case.

There are some clear trends worth noting. For the $\eta = 0.1$ case,
the $p$-state is an increasingly greater distance from the initial
position as time
increases. Over this time scale, or equivalently this hopping
distance, the small randomness does not strongly affect the motion
through the lattice. One interesting feature is the time scale
for motion. Although the nearest neighbor interaction energy
gives a time scale of 1.43~$\mu$s, the $p$ state has moved
$\simeq 14$ lattice spacings in 5~$\mu$s (approximately 3
scaled time units). Thus, the motion is somewhat faster than
might be expected.

For larger randomness, the hopping becomes increasingly restricted.
Comparing the $\eta = 0.1$ and $\eta = 0.2$ cases, it appears that
the farthest extent of the hopping is approximately the same
(about 14 sites) but the probability to be in the furthest peak
is $\sim 1/3$ as much for the larger randomness. For the
$\eta = 0.3$ and 0.4 cases, it appears that the distribution hardly
evolves for later times which indicates the $p$-state is restricted
to the region near where it started with the range decreasing
with increasing $\eta$.

\begin{figure}
\resizebox{80mm}{!}{\includegraphics{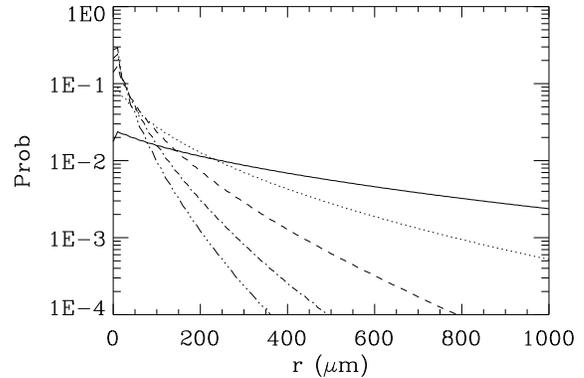}}
\caption{The asymptotic in time
probability for the $p$-state to be a distance
$r$ from its starting point averaged over a radial region
equal to a lattice spacing. All of these calculations are for
one dimension and have
Type (1) randomness with solid line corresponding to
$\eta = 0.1$, the dotted line to $\eta = 0.2$, the dashed line
to $\eta = 0.3$, the dash-dot line to $\eta = 0.4$ and
the dash-dot-dot-dot line corresponding to $\eta=0.5$.
These distributions are fit by the stretched exponential
decay $\exp (-[r/r_c]^\alpha)$ with $\alpha = 0.57\pm 0.02$
and $r_c$ decreasing with increasing randomness.
}
\end{figure}

An interesting question is whether the motion of the $p$-state is
actually restricted or whether its movement is only slowed down.
To address this, we can use Fig.~2 to show that the range is
restricted. This figure shows the asymptotic in time probabilities
for different amounts of Type (1) randomness (the asymptotic
probability to hop from site $i$ to site $j$ is given in Eq.~(\ref{eqAsym})).
All of the distributions have two kinds of decays. The initial,
fast decrease (to probabilities of $\sim 10^{-6}$) has the form of
a stretched exponential.
We fit these distributions using a simple function of the
form $C\exp (-[r/r_c]^\alpha )$ down to the values of the probability
of $10^{-10}$ or out to distances of $10^4$~$\mu$m which is 1000
sites. We found that $\alpha = 0.57\pm 0.02$ for all cases. The
``localization length scale", $r_c$, decreases with increasing
randomness approximately as $1/\eta^2$. Our fit values are
229~$\mu m$ for $\eta = 0.1$, 57.2~$\mu$m for $\eta = 0.2$,
22.8~$\mu$m for $\eta = 0.3$, 12.9~$\mu$m for $\eta = 0.4$,
and 9.5~$\mu$m for $\eta = 0.5$.
The slow decay seems to be a power law although the power could
not be accurately found by fitting because the data was noisy
at these low probabilities. A decrease like $1/r^6$ or $1/r^7$ is consistent
with the data.
The fact that the data has a power law decrease at large $r$
is due to the long range nature of the interaction between the
different basis functions.

For the $\eta = 0.1$ case, we performed a calculation with 16,000
sites. If there is one extended state, then the smallest probability
for the asymptotic distribution would be $\sim 2/16,000^2\sim 8\times 10^{-9}$.
In our calculation, the smallest probability was $\sim 10^{-12}$.
This means there are {\it no} extended states for this level
of randomness; thus, {\it all} states have a restricted range as would be
expected for Anderson localization.

\subsubsection{Type (2) Randomness}

\begin{figure}
\resizebox{120mm}{!}{\includegraphics{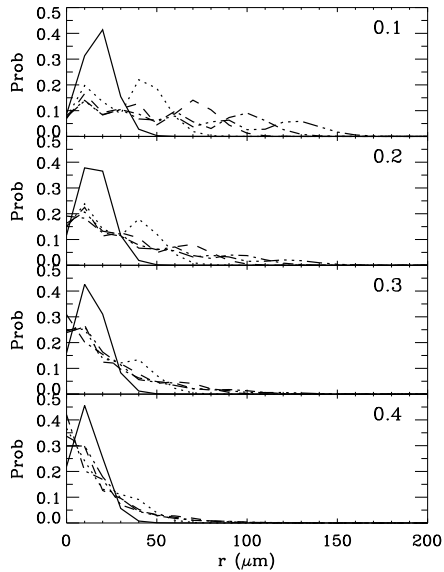}}
\caption{Same as Fig.~1 but for Type (2) randomness. The
plots are for different vacancy fractions $\zeta$.
}
\end{figure}

Figure~3 shows how the $p$-state hops away from its initial position
for different fractions of randomly missing atoms.
As with Fig.~1, the plots show the distribution
of hopping distances, Eq.~(\ref{eqDhop}), times the lattice
spacing, 10~$\mu$m, at different times;
the different lines correspond to different amounts of hopping
times (1, 2, 3, 4 and 5~$\mu$s).

There is a similar behavior to
that in Fig.~1. The case of least randomness has the maximum
extent of the hopping (approximately 14 sites) nearly the same as
the case of no randomness. However, there is less probability to reach
the furthest extent. As with Fig.~1, the hopping becomes increasingly
restricted with increasing randomness. Also, the $p$-state seems
to have reached the limit of its hopping range by $\sim 5$~$\mu$s
for the most random cases. Comparing the results from the two
types of randomness and setting $\eta =\zeta$,
it appears that missing sites have a larger effect on the motion.
For example, having 1/10 of the atoms missing (i.e., $\zeta = 0.1$) has
a larger effect than having all of the atoms randomly moved by a
distance of $\delta x/10$ (i.e., $\eta = 0.1$).

\begin{figure}
\resizebox{80mm}{!}{\includegraphics{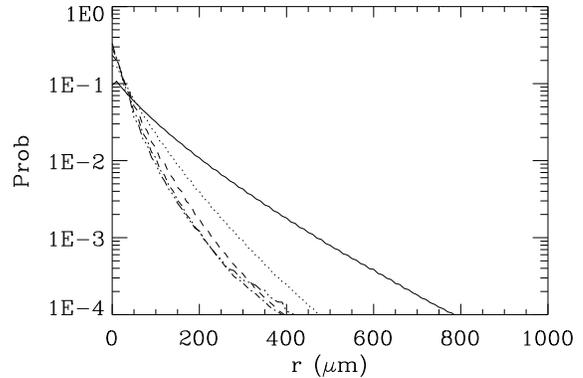}}
\caption{Same as Fig.~2 but for Type (2) randomness. The solid
line corresponds to
$\zeta = 0.1$, the dotted line to $\zeta = 0.2$, the dashed line
to $\zeta = 0.3$, the dash-dot line to $\zeta = 0.4$ and
the dash-dot-dot-dot line corresponding to $\zeta=0.5$.
These distributions are fit by the stretched exponential
decay $\exp (-[r/r_c]^\alpha)$ plus a power law
with $\alpha = 0.78\pm 0.02$
and $r_c$ decreasing with increasing randomness.
}
\end{figure}

As with Fig.~2, we can investigate whether the range of hopping is
actually restricted by examining the asymptotic in time distribution.
These results are shown in Fig.~4 and show a faster decrease
with distance compared to Fig.~2. The distribution becomes noisy
for probabilities less than $\sim 10^{-7}$. As with Fig.~2, there is
a fast decrease followed by a more slowly decreasing tail for
probabilities less than $\sim 10^{-6}$. The more slowly decreasing
part of the distribution was more prominent than for Type (1) noise
so we were able to fit both the fast and slow decay parts of the
distribution. We again found that the fast decay had the form
of a stretched exponential while the slow decay had the form of high power.
Because the slow decay was small there was a range of powers that
seemed to work nearly as well but a $1/r^7$ seemed to do best.
The form we fit to was
$C\exp (-[r/r_c]^\alpha )+B/(r+1.5\times 10^{-3} m)^7$. We found
that $\alpha = 0.78$ worked best for all cases with a spread of
$\pm 0.03$.
Our fit values are
64~$\mu m$ for $\zeta = 0.1$, 35~$\mu$m for $\zeta = 0.2$,
26~$\mu$m for $\zeta = 0.3$, 23~$\mu$m for $\zeta = 0.4$,
and 21~$\mu$m for $\zeta = 0.5$.

Comparing Figs.~2 and 4, it's clear that the Type (2) randomness
leads to a smaller range of hopping if we take $\sim 10^{-4}$ as
the condition. However, the localization lengths, which give
the $1/e$ condition, can be smaller or larger depending on the
amount of randomness. The reason for the difference in interpretation
is the larger
power in the stretched exponential for Type (2) randomness.

For the $\zeta = 0.1$ case, we performed a calculation with 8,000
sites. If there is one extended state, then the lowest probability
for the asymptotic distribution would be $\sim 2/8,000^2\sim 3\times 10^{-8}$.
In our calculation, the smallest probability was $\sim 10^{-14}$.
This means there are {\it no} extended states for this level
of randomness; thus, as with the Type (1)
randomness, {\it all} states have a restricted range as would be
expected for Anderson localization.

\subsection{Two dimensions}

In this section, we present the results of our calculations for
a two dimensional, square lattice. For the Type (1) randomness, the atoms
are only shifted in the $x$- and $y$-directions. The two
dimensional calculations are difficult to converge because
the number of atoms increases quadratically with the linear
lattice dimension. The time dependent calculations shown in
Fig.~5 are converged with respect to number of lattice
sites. None of the asymptotic distributions are converged with
respect to lattice size:
even the calculations with the most randomness have a large fraction
of extended states that cover the whole lattice.

For the two dimensional case, we only show the result for the
Type (1) randomness. While the randomness from missing sites
gave different results compared to the randomness from
shifting position, we did not find a qualitative difference
worth reporting.

\subsubsection{Type (1) Randomness}

\begin{figure}
\resizebox{120mm}{!}{\includegraphics{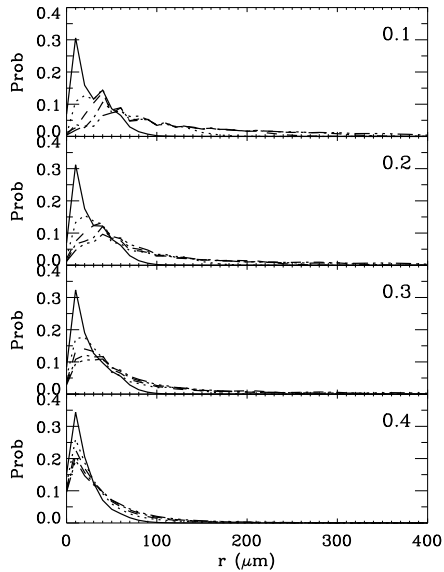}}
\caption{Same as Fig.~1 but for a two-dimensional lattice.
The results are for a $200\times 200$ lattice; thus, all of
these results are converged. Note, the different shape of
the distribution and the farther extent
compared to Fig.~1 which is the result of the
$p$-state having more atoms to interact with.
}
\end{figure}

Figure~5 shows the distribution of hopping distances at different
times for four levels of randomness. As in Figs.~1 and 3, the least random
case, $\eta = 0.1$, is very similar to the no randomness case.
The $\eta = 0.1$ and 0.2 cases qualitatively differ from the
same cases in one dimension (Fig.~1). For Fig.~5, the peak in the
distribution is at smaller $r$, but the 5~$\mu$s
distribution noticeably differs from 0 for a range more than
twice that in Fig.~1. These differences are a reflection of the
different band structure for two dimensions compared to one
dimension (Figs.~2 and 3 of Ref.~\cite{RHT}). The $\eta = 0.3$
and 0.4 cases do not appear to be qualitatively different from
the corresponding cases in Fig.~1. They both appear to have reached
their maximum extent by approximately 5~$\mu$s. The $\eta=0.3$
case has a larger extent than the 0.4 case which is expected
since larger randomness should more strongly confine the $p$-excitation.

\begin{figure}
\resizebox{80mm}{!}{\includegraphics{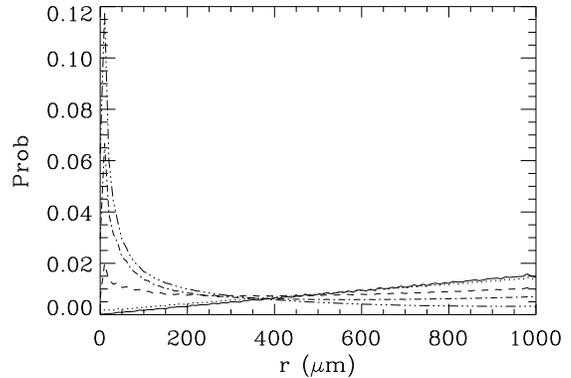}}
\caption{Same as Fig.~1 but for a two dimensional lattice.
The results are for a $200\times 200$ lattice. Since the
wrap condition starts at 1000~$\mu$m and {\it all} cases
have probability out to that distance, {\it none} of the
calculations are converged.
The linear increase with $r$ for small $\eta$ is due to the
linear increase in number of lattice sites with $r$. The
peak at small $r$ for larger $\eta$ is from a fraction of
localized states.
}
\end{figure}

As with Figs.~2 and 4, we can investigate whether the hopping is
slowed by the disorder or is stopped by plotting the asymptotic
in time distribution of hopping distances. Figure~6 shows this
for the five different levels of Type (1) randomness for a lattice
of $200\times 200$ atoms. Notice that Fig.~6 has a linear $y$-axis
whereas Figs.~2 and 4 have a log-scale. Unlike the one dimensional cases in
Figs.~2 and 4, the probability extends to the edge of the lattice
for all cases. Thus, none of these curves are fully converged.
The case with the least randomness, $\eta = 0.1$, is nearly
indistinguishable from the no randomness calculation; the probability
increases linearly with distance because the number of sites between
$r$ and $r+10\mu$m increases linearly with distance. This means
nearly all states for $\eta = 0.1$ extend for linear distance
of over 100 lattice sites (i.e. nearly all states cover
$\sim 10^4$ or more sites). Compare this with the
$\eta = 0.1$ line in Fig.~2 which has an order of magnitude decrease
in hopping probability over the same hopping range.
The $\eta = 0.2$ case only slightly
differs from the $\eta = 0.0$ case with slightly higher probability
at smaller $r$ and slightly lower probability at larger $r$; in
Fig.~2, the $\eta = 0.2$ case had a decrease in probability of more
than a factor of 100 over the range shown.

The cases with large
randomness show a peak at small $r$ which reflects the existence
of localized states. Since the localized states do not extend to
the edge of the lattice, the small $r$ behavior is nearly converged
for $\eta \geq 0.3$. Roughly, the region of convergence is $\leq 50$~$\mu$m
for $\eta = 0.3$ and $\leq 400$~$\mu$m for $\eta = 0.5$. As with
the smaller randomness cases, the localization region is much larger
than that for the corresponding one dimensional cases.

A final difference
between the one and two dimensional calculation is how the results
change with increasing randomness. The one dimensional case had
a smooth change in the asymptotic properties with increasing randomness.
The two dimensional case has almost no localized states for $\eta = 0.1$
and 0.2 with a big jump in number of localized states when
going from $\eta = 0.2$ to 0.3.

\subsection{Three dimensions}

\begin{figure}
\resizebox{120mm}{!}{\includegraphics{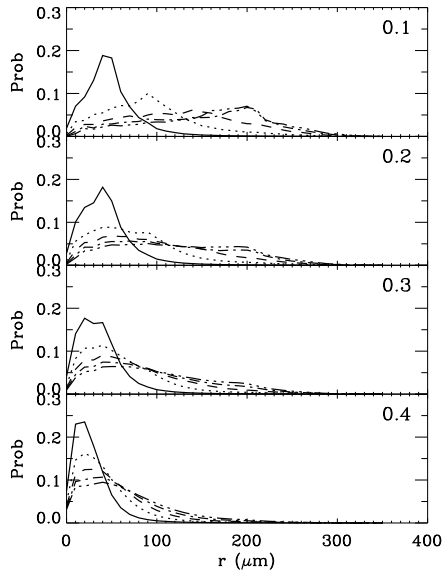}}
\caption{Same as Fig.~1 but for a three-dimensional lattice.
The results are for a $40\times 40\times 40$ lattice. Since the
wrap condition starts at 200~$\mu$m, any of the probability
distributions that extend past this are not converged.
Note, the different shape of
the distribution and the farther extent
compared to Figs.~1 and 5 which is the result of the
$p$-state having more atoms to interact with.
}
\end{figure}

In this section, we present the results of our calculations for
a three dimensional, cubic lattice. The three
dimensional calculations are the most difficult to converge because
the number of atoms increase cubically with the linear
lattice dimension. The largest calculation we performed was for
a lattice of $40\times 40\times 40$ atoms (i.e. 64,000 total).
The wrap boundary condition starts for atoms differing by 20 lattice
sites in any direction. This means only the distances less than
200~$\mu$m do not depend on the wrap condition.
The time dependent calculations shown in
Fig.~7 are not converged with respect to number of lattice
sites; the most nearly converged is the $\eta = 0.5$ case since,
for that case,
there was only a small probability to hop more than 20 sites
during the time shown.
None of the asymptotic distributions are converged with
respect to lattice size:
even the calculations with the most randomness mostly consist
of extended states that cover the whole lattice.

As with the two dimensional case, we only show the Type (1) randomness
because the results of randomly removing atoms from sites are
similar in character.

\subsubsection{Type (1) Randomness}

Figure~7 shows the time dependence of the distribution of hopping
distance for four different levels of randomness. For the later
times, only the $\eta = 0.4$ case is converged. The other calculations
show a distinct change in slope at $r=200$~$\mu$m. This is the distance
corresponding to the wrap boundary condition and is an artifact.
Despite the lack of convergence, there is some useful information
that can be extracted. For example, it is clear that the $p$-excitation
hops even further than the two-dimensional case, Fig.~5. Thus, the
extra interactions that arise in higher dimension increase the
speed of the hopping. Another example is the relatively small effect that the
randomness has. The $\eta = 0.1$ and $\eta = 0.2$ hopping distributions
are quite similar. Also, the $\eta = 0.4$ case is still has a clearly
evolving hopping distribution at 5~$\mu$s unlike the one- and
two-dimensional cases.

\begin{figure}
\resizebox{80mm}{!}{\includegraphics{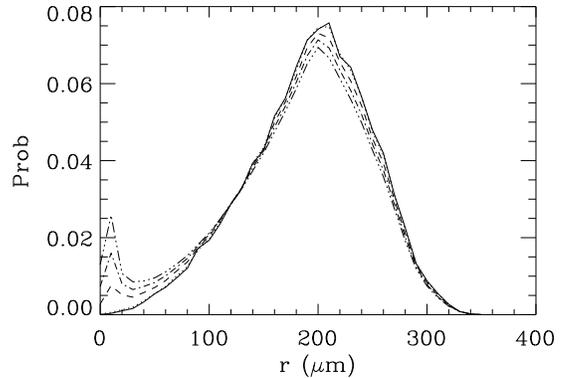}}
\caption{Same as Fig.~1 but for a three dimensional lattice.
The results are for a $40\times 40\times 40$ lattice. Since the
wrap condition starts at 200~$\mu$m and {\it all} cases
have probability out to that distance, {\it none} of the
calculations are converged.
The quadratic increase with $r$ is due to the
quadratic increase in number of lattice sites with $r$ out to
200~$\mu$m; the decrease for $r>200$~$\mu$m is because the
sphere extends outside of the cube in our calculation. The
peak at small $r$ for larger $\eta$ is from a fraction of
localized states.
}
\end{figure}

Figure~8 gives the asymptotic distribution of hopping distances
for different $\eta$. The $\eta = 0.1$ and 0.2 cases hardly differ
from the no randomness case. These distributions simply reflect the
number of sites versus distance. For $r\leq 200$~$\mu$m, the number
of sites between $r$ and $r+\delta x$ increases quadratically with $r$.
For larger $r$ the number of sites decreases because the wrap-cube
is only filled out to 200~$\mu$m. For $r\geq 400$~$\mu$m there are
no sites. The $\eta = 0.3$, 0.4 and 0.5 cases have a small peak at
$r = 10$~$\mu$m which arises from a small fraction of localized
states. If one counts the extra probability for $r \leq 60$~$\mu$m,
there is less than 10\%
of the states localized even for $\eta = 0.5$. Thus, we expect that
almost all excitations will be delocalized in three dimensions unless
the randomness is even greater than the cases we considered.

\section{Conclusions}

We have performed calculations for how a $p$-state hops through
a sea of $s$-states due to the dipole-dipole interaction.
We focussed on how the hopping changes when the atoms are
positioned on a perfect lattice or have randomness. We considered
two kinds of randomness: (1) the atoms have a slight, random
shift from a position and (2) random atoms are removed from
a perfect lattice. The case of a one-dimensional lattice
gave the largest qualitative difference between the two kinds
of randomness.

Our one dimensional calculations with randomness resulted
in all of the states being localized independent of the type 
of randomness or the size of randomness. The distribution
of hopping distances could be fit with a stretched exponential
whose exponent depended on the type of randomness but did
not depend on the size of the randomness. This suggests that
even miniscule randomness would lead to all states being
localized.
It is not surprising that the one dimensional cases with 
randomness lead to localization even for small amounts of
randomness. However, the long range interaction in the Hamiltonian
leads to a power law decrease with hopping distance.

For two- and three-dimensions, it appears that the randomness
slows down the hopping but leads to localized states only
for large randomness. It appears that the number of localized
states goes to 0 as the randomness decreases; the number of localized
states might be 0 even for small, but non-zero, randomness.

This work was supported by the National Science Foundation.


\end{document}